\documentstyle[aps,twocolumn,epsf]{revtex} %%\tightenlines

\def\lsim{\lower.8ex\hbox{$\buildrel<\over\sim$}}
\def\gsim{\lower.8ex\hbox{$\buildrel>\over\sim$}}
\begin{document} 
\draft
\twocolumn[\hsize\textwidth\columnwidth\hsize\csname
@twocolumnfalse\endcsname
%%\tightenlines
\title {Discrete solvent effects on the effective 
interaction between charged colloids}
%%\bigskip
\author{E. Allahyarov  and H. L{\"o}wen}

%%\bigskip\bigskip
\address{Institut f{\"u}r Theoretische Physik II, 
Heinrich-Heine-Universit{\"a}t D{\"u}sseldorf, D-40225 D{\"u}sseldorf,
Germany}
%\narrowtext
\date{\today}

\maketitle
\begin{abstract}

 Using computer simulations of 
two charged colloidal spheres with their counterions in a hard sphere solvent,
we show that  the granular nature of the solvent significantly influences
 the effective colloidal interaction.
For divalent counterions, the total effective 
force can become attractive generated by counterion hydration, while
for  monovalent counterions the forces are repulsive and well-described  by a 
solvent-induced  colloidal charge renormalization. Both effects are 
not contained  in the traditional ``primitive" approaches but can be accounted for 
in a solvent-averaged primitive model.
\end{abstract}
%%\vskip2pc]
]
\renewcommand{\thepage}{\hskip 8.9cm \arabic{page} \hfill Typeset
using REV\TeX }
\narrowtext

Supramolecular aggregates, such as colloids, polymers  or biological macromolecules,
are typically suspended in a molecular solvent 
which guarantees their stability and profoundly influences their
viscoelastic properties \cite{ref1}: 
examples range from paints to  dense DNA-solutions in biological cells. 
A full ``ab initio" statistical description of supramolecular solutions
should start from a molecular scale including the solvent explicitly. 
Obviously this is a tremendous task due to the large length scale separation
between the microscopic solvent  and the mesoscopic solute
and the enormous number of solvent particles which have to be considered explicitly. Therefore,
most of the common statistical approaches are based on so-called ``primitive" 
models  insofar as they disregard the molecular nature of the solvent completely such that
solvent properties only enter via a continuous 
background.

A particular example for  such a separation of length scales are charged
colloidal suspensions \cite{HansenLoewen} consisting of
highly charged mesoscopic particles (so-called polyions) suspended in water
or any other organic solvent together with their oppositely charged microscopic counterions.
The key quantity to understand the stability, structure and dynamics of such colloidal
dispersions is the {\it effective interaction\/} between two polyions, as a 
function of their mutual distance $r$. Neglecting 
 the discrete solvent this quantity has been calculated 
using  modeling on different descending levels: i)
the ``primitive model" (PM) of strongly asymmetric electrolytes
which takes into account explicitly the counterions 
ii) the nonlinear Poisson-Boltzmann approach which is inferior
to the PM as it neglects counterion correlations, iii) 
the linearized screening theory resulting in a Yukawa form for the
effective interaction potential as
given by the electrostatic part of the celebrated
Derjaguin-Landau-Verwey-Overbeek (DLVO) theory \cite{Verwey}.
The main effects  of nonlinear Poisson-Boltzmann theory 
can be encaptured by a similar Yukawa potential but with  ``renormalized" parameters
leading to the concept of colloidal charge renormalization 
\cite{Alexander}.
This picture is consistent with experimental data in dilute  bulk solutions
with monovalent counterions \cite{exp1,exp2}. 
Very strong correlations between  divalent and trivalent counterions, however,
may  lead to  attractive effective forces
between like-charge polyions as shown in recent computer simulations of the PM
\cite{AllahyarovPRL,Linse,Messina}.
\begin{figure}
   \epsfxsize=7.8cm %6cm
   \epsfysize=7.8cm%7cm
  ~\hfill\epsfbox{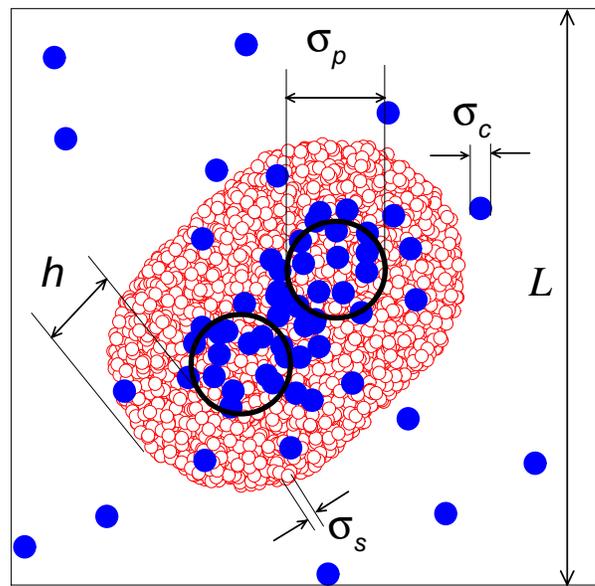}\hfill~
   \caption{View of the set-up as a projection of a simulation snapshot:
 Two polyions (dark open circles) in a bath of solvent particles (small hollow spheres)
contained in a spherocylindric cell  of width $h$.
The counterions  shown as small dark spheres can move in the whole simulation box of size $L$.}
     \label{fig_1}
\end{figure}

In this Letter, we investigate the influence
 of {\it solvent granularity\/}  on the effective interactions
between charged colloids. We explicitly add to the PM  a molecular
solvent modelled by a hard sphere fluid. We study this model  by direct
computer simulation and use the concept of effective interactions
to overbridge the gap between microscopic and mesoscopic
length scales. Our motivation to do so is twofold:
First, although the dipolar nature of the solvent
 \cite{LIE2} is not included, 
the model provides a minimal framework towards a statistical description
of hydration forces. Second,
the solvent hard sphere model was considered  in earlier studies for the effective
interaction between charged plates 
using liquid integral equations \cite{LIE}, modified Poisson-Boltzmann
 theory \cite{MPB} 
or more sophisticated density functional approaches \cite{DFT}. 
All these descriptions, however,
suffer from additional uncontrolled approximations such that  ``exact" computer 
simulation results are highly desirable. Such simulations were performed for parallel
plates \cite{Henderson2} and for small neutral particles  \cite{CS}
 but are hitherto not available for spherical charged colloids.

We implement a new ``solvent bath" simulation scheme which allows to simulate many neutral spheres
together with the charged species and obtain explicit results for the effective
force between nano-sized highly charged  colloids. We use these data to test a theory
with solvent-averaged effective interactions between the charged particles similar
in spirit to the old McMillan-Mayer approach for electrolyte solutions \cite{MM}. This 
solvent-averaged  primitive model (SPM) yields good agreement with the simulation data
and can thus be used  to obtain the effective interaction
between larger colloidal particles.
For monovalent counterions and large distances $r$, 
 our simulation data can be described perfectly  within a Yukawa-potential
with a solvent-induced  polyion charge 
and screening length renormalization. For divalent counterions and nano-sized colloids,
we find an attractive force. Both effects are not contained in the PM. 

We consider two large spherical polyions  with diameter $\sigma_p$ and
charge $q_p$ 
at  distance $r$, together with their  counterions of diameter
$\sigma_c$ and charge $q_c$ 
in a bath of a neutral solvent $(q_s=0)$ with diameter $\sigma_s$. In our model,
the pair potentials
between the particles as a function of the mutual distances $r$ 
are  a combination of excluded volume and Coulomb terms
$$
V_{ij} (r) = \cases { \infty &for $ r \leq (\sigma_i + \sigma_j)/2 $\cr
   {{q_iq_j} / {\epsilon r}} &else\cr}
\eqno(1)
$$
where $\epsilon$ is the a smeared background dielectric constant of the solvent
and $(ij)=(pp),(pc),(ps),(cc),(cs),(ss)$.
Further parameters are the thermal energy $k_BT$ and the partial number densities
$\rho_i \ \ (i=p,c,s)$ which can be expressed as partial volume fractions
$\phi_i= \pi \rho_i \sigma_i^3/6 \ (i=p,c,s)$.
Charge neutrality requires $\rho_p | q_p |= \rho_c | q_c |$.
We fix the two polyions along the body diagonal in a cubic simulation box of length $L$
with periodic boundary conditions, hence $\rho_p=2/L^3$.
In a dense  fluid solvent ($\phi_s\approx 0.3$),
 many solvent spheres are in the box, such that a direct simulation is
 impossible. Thus we resort to the following ``solvent-bath" 
procedure: we define a spherocylindrical cell  around the colloidal pair  such that the minimal
distance $h$ from the colloidal surface to the cell boundary is  larger
than the  hard sphere bulk correlation length $\xi$,
see Figure 1 for the  set-up. The hard sphere solvent is only contained in this spherocylinder
 while the counterions can move within 
the whole simulation box. 
 We use the Molecular Dynamics (MD)
method to calculate the particle trajectories. 
Once a solvent particle is leaving the spherocylindrical cell, it is
randomly inserted at another  
place of the cell boundary with the same velocity and a random depth in
order to avoid unphysical solvent layering on the cell surface.   
 Since  $h$ is much larger than $\xi$, 
the presence of the boundary has no influence on the inhomogeneous
density distribution of the solvent 
and the counterions near the colloidal surfaces.
For  a  counterion approaching  the cell boundary, however, there is
an artificial  asymmetry between the solvent bath inside the cell and 
the ``vacuum" outside the cell which hinders a counterion to penetrate 
into the solvent bath. This unphysical effect is repaired in the simulation scheme
by switching off the counterion-solvent interaction for a counterion which
is penetrating from outside. Once the counterion is fully
surrounded by solvent molecules the interaction is turned on again. 
This procedure guarantees a symmetric crossing rate of counterions
across the spherocylindrical cell. 
In the solvent-free space outside the cell, the counterion-counterion interaction
is still $V_{cc}(r)$ as the mean counterion distance is much larger than $\xi$
such that the Coulomb repulsion dominates solvent depletion effects.
The algorithm was carefully tested 
for solvent slabs between charged plates and perfect agreement was found compared to
simulations where the whole space was filled with solvent particles.

In our simulations, we fixed   $T=298^oK$ and 
$\epsilon=81$ (water at room temperature) with $\sigma_s=3\AA$, $\phi_s=0.3$
(such that $\xi$ is about $3\sigma_s$) and $\sigma_c=6\AA$. The width
$h$ is $12\sigma_{s}$  
such that typically
$N_s=25.000-30.000$  solvent hard spheres are simulated. 
We varied the polyion size $\sigma_p$ 
and charge $q_p$  and calculated the solvent- and counterion-averaged 
total force acting onto a polyion for a given colloidal distance $r$. The force is projected onto the
separation vector of the two colloidal spheres such that a positive sign means repulsion.
This effective  force $F(r)$ is the sum of four different contributions \cite{AllahyarovPRL}:
 the direct Coulomb repulsion
as embodied in $V_{pp}(r)$, the counterion screening resulting from the averaged
Coulomb force of counterions acting onto the polyions, 
the counterion depletion term arising from the 
hard sphere part of $V_{pc}(r)$,
and the solvent depletion force. 

For nano-sized colloids, explicit results for  $F(r)$  are presented
in Figures 2a and 2b. For nearly touching polyions (full curves in the
insets of Figs. 2a and 2b)
the force exhibits oscillations on the scale of the solvent diameter due to
solvent layering leading to attraction for touching polyions
as the attractive solvent depletion part exceeds the bare Coulomb repulsion. For larger distances
and  monovalent counterions, the force is repulsive. Simulation data for the
PM are also included which overestimate the force. The repulsion is even stronger
in DLVO theory  as derived from the Yukawa pair potential
$$
V(r)= {{q_p^2\exp ( -\kappa (r-\sigma_p ))}\over{(1+\kappa\sigma_p/2)^2 \epsilon r }} \eqno(2)
$$
with $\kappa=\sqrt{4\pi\rho_c q_c^2/\epsilon k_BT}$. 
For divalent counterions, on the other hand,
 there is attraction within a range of several polyion
diameters. This attraction originates from counterion overscreening
induced by hydration, as the pure PM yields repulsive forces.

Intermediate between the PM and the full solvent simulation, we put forward
a description on the primitive 
\twocolumn[\hsize\textwidth\columnwidth\hsize\csname
@twocolumnfalse\endcsname
\begin{figure}
   \epsfxsize=7.8cm %6cm
   \epsfysize=7.5cm %7cm
~\hfill\epsfbox{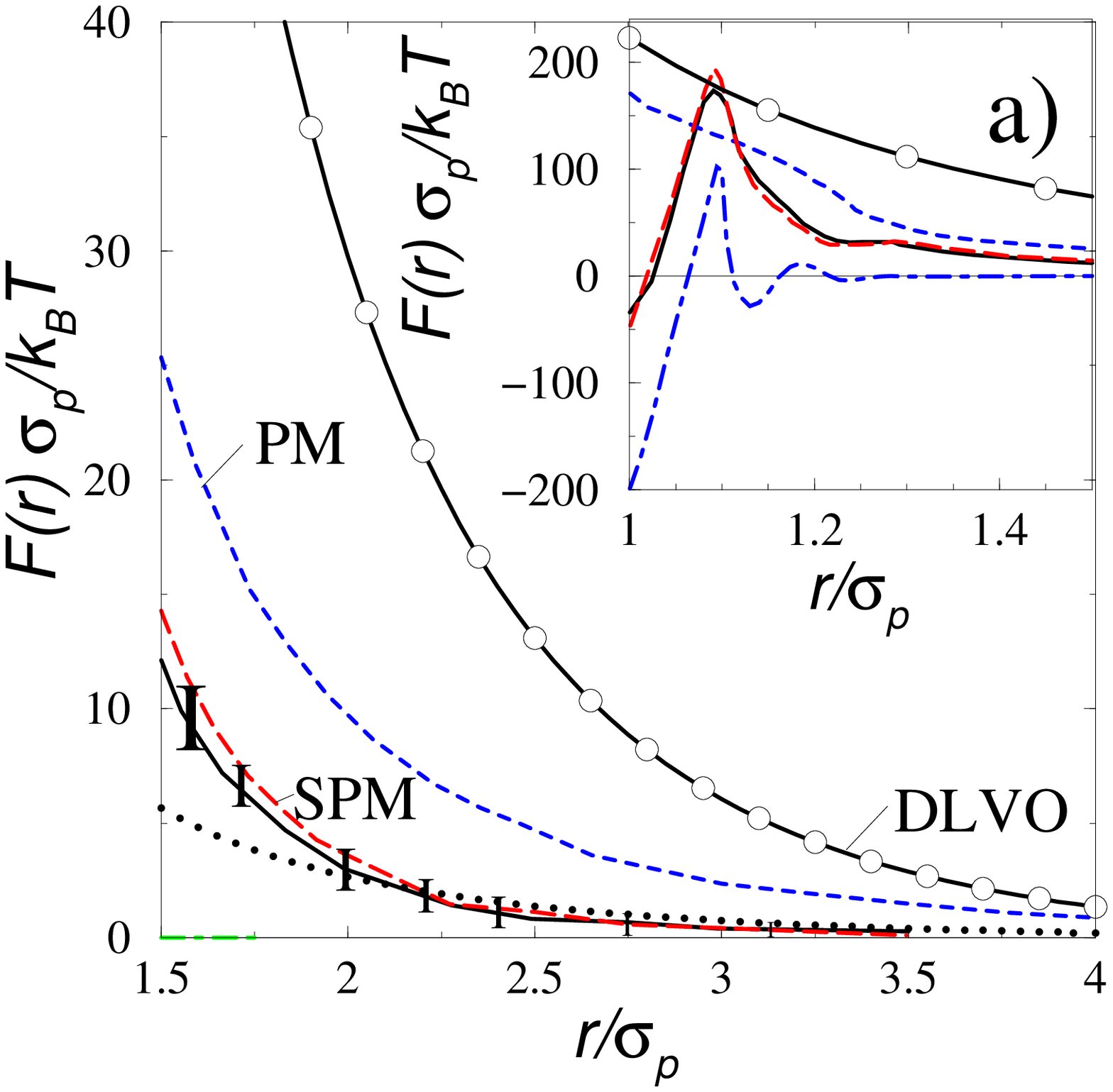}\hfill~
   \epsfxsize=7.8cm %6cm
   \epsfysize=7.5cm%7cm
~\hfill\epsfbox{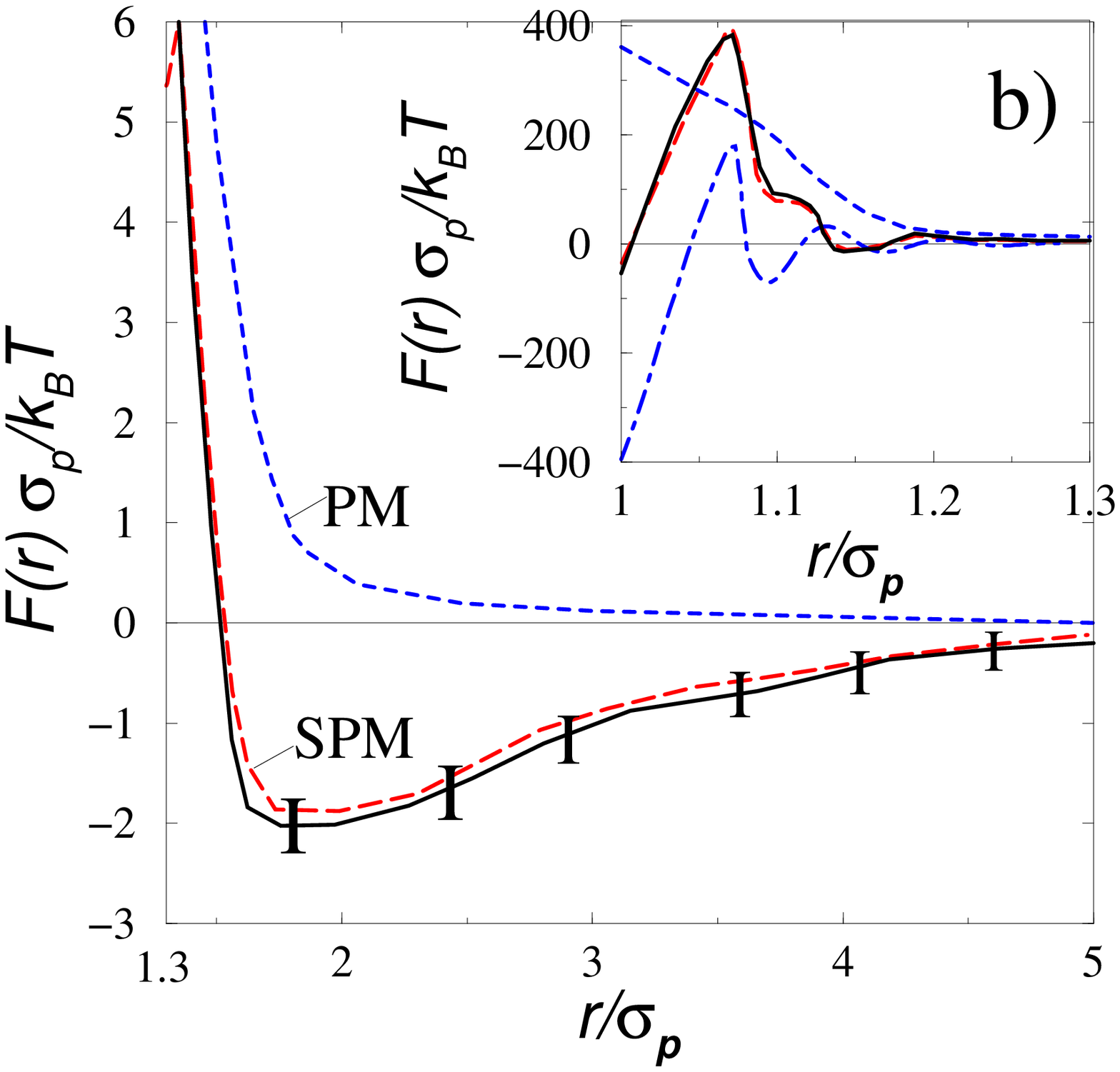}\hfill~
 \caption{Reduced distance-resolved force $F(r)\sigma_p /k_BT$ versus reduced
distance $r/\sigma_p$. The inset shows the same for nearly touching
polyions of molecular distances.\
a $)$ for monovalent counterions and  $\sigma_p : \sigma_c : \sigma_s
= 10:2:1$; b$)$ for divalent counterions and  
$\sigma_p : \sigma_c : \sigma_s = 14:2:1$. The further parameters are  
$\vert q_p/q_c \vert =32$ and  $\phi_p=5.8\times 10^{-3}$. Solid line with error bars:
 full solvent bath simulation; long-dashed line: SPM; short-dashed
 line: PM;
open circles: DLVO theory; dot-dashed line in inset: solvent depletion
force. The dotted line in a$)$ 
is the solvent-renormalized Yukawa model.}
    \label{fig_2}
\end{figure}
] 
\hspace{-0.6cm} level {\it with  solvent-averaged interactions\/} between the
charged particles which provides  a simple physical picture of 
discrete solvent effects. In fact, by tracing out the solvent degrees of freedom exactly,
one arrives at additional solvent-induced  interactions on the primitive level
for which we approximately only consider pairwise terms. This  defines the 
solvent-averaged primitive model (SPM). The additional depletion
interactions in hard sphere mixtures have been 
studied extensively by theory \cite{depl1} and simulation \cite{depl2}.
The polyion-polyion solvent-induced depletion force is also included in Figure 2
exhibiting oscillations on the scale of the molecular solvent diameter
$\sigma_s$. 
While the  additional solvent-averaged counterion-counterion
interaction is  much smaller 
than the Coulomb repulsion, the solvent-averaged  polyion-counterion
interaction  
  results in  a deep counterion attraction 
towards  the colloidal surfaces with a molecular range and 
a potential energy depth of several $k_BT$, describing granular
hydration forces.  
 The only approximation used to derive the SPM is that
solvent-induced many-body forces between the charged particles are ignored.
This is justified as typical distances between triplets of charged
particles are larger than 
$\xi$ except for nearly touching polyions with
``squeezed"  counterions. We determine the additional depletion
interaction of the SPM by a reference simulation of two spheres in a
hard sphere solvent \cite{depl2} and use them as an input for a
simulation of the SPM\@. As can be deduced from Figure 2,
the SPM  describes the solvent bath data extremely well, yielding
results that lie within the statistical error of the full simulation
over the whole range of distances.

We finally use the SPM to investigate solvent effects for
polyion sizes in the colloidal domain.  
Distance-resolved colloidal forces $F(r)$ for monovalent counterions 
are presented in Figure 3. These forces
 are repulsive but  much smaller than those from PM simulations or
 DLVO theory.
This is due to counterion accumulation near the colloidal surface as induced by the 
additional solvent depletion attraction. As 
 the corresponding potential energy gain is only few $k_BT$, this depletion attraction is 
different from chemisorption of counterions. 
For very large distances, on the other hand, the screening of the
remaining free counterions will dominate the interaction which can be 
described by a  cell model \cite{Alexander}. To test this, we have performed additional 
solvent-bath simulations for a single polyion in a spherical cell of radius 
$R=(4\pi\rho_p /3)^{-1/3}$ calculating the counterion 
density ${\tilde \rho}_c$ at the cell boundary. The corresponding effective Yukawa potential 
 has the same form as in eq.(2) but with a solvent-renormalized screening length 
$\kappa^* = \kappa \sqrt{{\tilde \rho}_c /\rho_{c }} $
and a solvent-renormalized  charge $q_p^*=q_p {\tilde \rho}_c /\rho_c$
which is considerably smaller than the bare charge. 
The actual value of this  renormalized charge, however, differs strongly
 from the charge renormalization according
to the PM or Poisson Boltzmann theory \cite{Alexander}.
The  force resulting from the solvent-renormalized Yukawa model
fits our full simulation data of nano-sized colloids for large distances 
and monovalent counterions (see Figure 2a) and
 perfectly describes the SPM data for larger colloids
except for molecular  distances (see Figure 3). Consequently,
a Yukawa model can still be used but the parameters have to be suitably renormalized.
Clearly, the repulsive Yukawa model breaks down if no free counterions
are left as is the case for, {e.g.\/}, divalent counterions and nano-sized colloids. 
\begin{figure}
   \epsfxsize=7.8cm %6cm
   \epsfysize=7.3cm %7cm
~\hfill\epsfbox{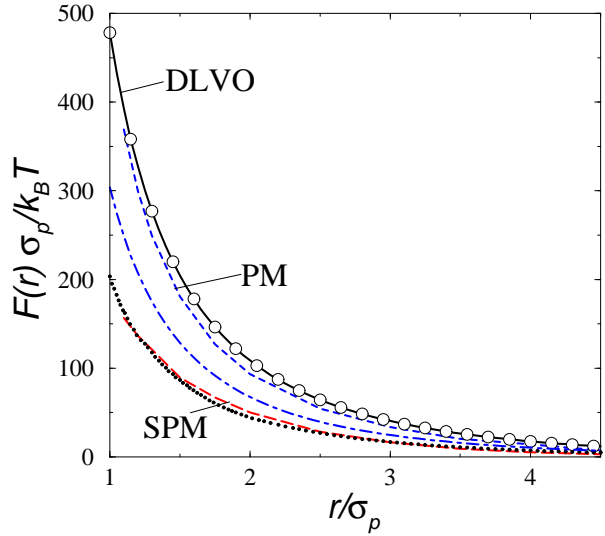}\hfill~
   \caption{Same as Fig.2 but now for larger polyions
$\sigma_p:\sigma_c:\sigma_s = 370:2:1$, $\vert q_p/q_c \vert =280$, $\phi_p=2.3 \times 10^{-3}$, and
monovalent counterions. 
The dot-dashed line here is the PM with colloidal charge reduction by
the corresponding  
average amount of counterions in a shell of width $\xi$
around the colloids. The meaning of the other lines is as in
Figure 2a. }
    \label{fig_3}
\end{figure}
We finally remark that the traditional justification of the PM 
is to define a bare charge by fixing all the counterions which are in
a molecular shell around the colloids and to treat the rest within the
PM\@. This picture of polyion charge reduction  
 works for chemisorption but not for the
weaker hydration forces. In fact, if the polyion charge is reduced by the corresponding
average counterion charge in a shell of thickness of the solvent
bulk correlation length $\xi$, the PM still overestimates the force
considerably, see the dot-dashed line Figure 3.

In conclusion, based on a unified statistical description of 
counterion hydration and screening, we have shown that hydration
forces profoundly influence  
 the colloidal interaction. For divalent counterions, there is
 solvent-induced {\it attraction\/}  
 which is not contained in  the traditional primitive model but can be
 encaptured  
within a solvent-averaged primitive model.
For monovalent counterions, the forces 
 can be described by a  {\it charge renormalization\/}
induced by counterion  hydration forces.
This picture is in agreement  with experiments on strongly deionized samples
where a Yukawa picture can still be employed,
provided the colloidal charge is renormalized towards a value  smaller
than the bare charge \cite{exp1,Gisler}. The general concept of charge renormalization
should be  transferable to other situations where screening by few remaining ``free" counterions 
dominates the interaction as, {e.g.\/}, for
a polar solvent, for different polyionic surface charge patterns and for  explicit surface chemistry.

We  thank R. Roth, C.N. Likos and T. Palberg for  helpful remarks.

%\bibitem{Palberg2} F. Bitzer, T. Palberg, H. L\"owen, R. Simon, P. Leiderer, Phys. Rev. E {\bf 50}, 2821 (1994).
%\bibitem{Durand} R. V. Durand, C. Franck, Phys. Rev. E {\bf 61}, 6922 (2000).
%\bibitem{planar} S. Marcelja, Colloids and Surfaces A {\bf 129-130} 321 (1997).
%big particles are neutral

\end{document}